\documentclass[aps,preprintnumbers,superscriptaddress,showpacs]{revtex4}
\usepackage{epsfig}
\usepackage{psfrag}
\usepackage{amsfonts}
\usepackage{graphicx}
\usepackage{dcolumn}
\usepackage{bm}

\begin{document}

\title{Drag force on heavy quarks from holographic QCD}

\author{Yuanhui Xiong}
\affiliation{School of Mathematics and Physics, China University
of Geosciences, Wuhan 430074, China}
\author{Xingzheng Tang}
\affiliation{School of Mathematics and Physics, China University
of Geosciences, Wuhan 430074, China}
\author{Zhongjie Luo}
\email{luozhj@cug.edu.cn} \affiliation{School of Mathematics and
Physics, China University of Geosciences, Wuhan 430074, China}

%%%%%%%%%%%%%%%%%%%%%%%%%%%%%%%%%%%%%%%%
\begin{abstract}
We study the drag force of a relativistic heavy quark using a
holographic QCD model with conformal invariance broken by a
background dilaton. We analyze the effects of chemical potential
and confining scale on this quantity, respectively. It is shown
that the drag force in this model is larger than that of
$\mathcal{N}=4$ supersymmetric Yang-Mills (SYM) plasma. In
particular, the inclusion of the chemical potential and confining
scale both enhance the drag force, in agreement with earlier
findings. Also, we discuss how chemical potential and confining
scale influence the diffusion coefficient.
\end{abstract}
\pacs{11.15.Tk, 11.25.Tq, 12.38.Mh}

\maketitle
%%%%%%%%%%%%%%%%%%%%%%%%%%%%%%%%%%%%%%%%
\section{Introduction}
The ultra-relativistic heavy-ion experimental programs at the
Relativistic Heavy Ion Collider (RHIC) and the Large Hadron
Collider (LHC) have created a new type of matter so-called quark
gluon plasma (QGP) \cite{EV,KA,JA}. One of the interesting
features of QGP is jet quenching (parton energy loss): high energy
partons propagate through the hot and dense medium, they interact
with the medium and hence lose energy. Usually, the jet quenching
phenomenon could be characterized by the jet quenching parameter,
defined by the average transverse momentum square transferred from
the traversing parton, per unit mean free path \cite{XN}.
Alternately, this phenomenon could be studied from the drag force:
heavy quarks move through the QGP, they feel a drag force and
hence lose energy. In weakly coupling theories, the jet quenching
has already been investigated in many works
\cite{BG0,RB2,RB1,UA,AMY,SJ,GLV}. However, a lot of experiments
find strong evidence that QGP does not behave as a weakly coupled
gas, but rather as a strongly coupled fluid. Consequently,
calculational techniques for strongly coupled, real-time QCD
dynamics are required. Such techniques are now available via the
gauge/gravatity duality or AdS/CFT correspondence
\cite{Maldacena:1997re,MadalcenaReview,Gubser:1998bc}.

AdS/CFT, a conjectured duality between a string theory in the AdS
space and a conformal field theory on the AdS space boundary, has
yielded lots of important insights for exploring various aspects
of QGP \cite{JCA}. In the frame work of AdS/CFT, the drag force of
a heavy quark moving in $\mathcal{N}=4$ SYM plasma was proposed in
\cite{GB,CP}. Therein, this force could be related to the damping
rate $u$ (or called friction coefficient), defined by the Langevin
equation, $dp/dt=-u p+f$, subject to a driving force $f$. For
$dp/dt=0$, i.e., constant speed trajectory, $f$ equals to the drag
force. It was found that \cite{CP,GB}
\begin{equation}
f_{SYM}=-\frac{\pi\sqrt{\lambda}T^2}{2}\frac{v}{\sqrt{1-v^2}},
\end{equation}
where $v$, $T$, $\lambda$ are the quark velocity, the plasma
temperature and 't Hooft coupling, respectively. Subsequently,
this idea has been generalized to various cases, such as chemical
potential \cite{ECA,SCH,LC}, finite coupling \cite{KBF},
non-commutativity plasma \cite{TMA} and AdS/QCD models
\cite{PE,ENA,UGU}. Other studies in this direction can be found in
\cite{OA0,ZQ,DGI,KLP,ANA,MCH,SRO,SSG1,JSA}.

In this paper we study the drag force in a soft wall-like model,
i.e., the SW$_{T,\mu}$ model, which is motivated by the soft wall
model of \cite{AKE}. The SW$_{T,\mu}$ model was proposed in
\cite{PCO} to investigate the heavy quark free energy and the QCD
phase diagram. It was shown that such a model provides a nice
phenomenological description of quark-antiquark interaction as
well as some hadronic properties. Recently, the authors of
\cite{zq} investigated the imaginary potential in this model and
found that the presence of the confining scale tends to reduce the
quarkonia dissociation, reverse to the effect of the chemical
potential. Also, there are other investigations of models of this
type \cite{OA1,CPA,PCO2,PCO1,XCH,CE}. Motivated by this, here we
consider the drag force in this model and understand the possible
physical implications of our results in this dual plasma. Another
motivation for this paper is that previous works either discuss
the drag force in some strongly coupled thermal gauge theories
(not holographic QCD) with chemical potential \cite{ECA,SCH,LC} or
holographic QCD without chemical potential \cite{ENA,PE,UGU}. Here
we give such analysis in holographic QCD with chemical potential.

The structure of this paper is as follows. In the next section, we
introduce the geometry of the SW$_{T,\mu}$ model given in
\cite{PCO}. In section 3 and section 4, we analyze the behavior of
the drag force and the diffusion coefficient for the SW$_{T,\mu}$
model, in turn. Summary and discussions are given in the final
section.

%%%%%%%%%%%%%%%%%%%%%%%%%%%%%%%%%%%%%%%%

\section{Setup}
To begin with, we introduce the holographic models in terms of the
action \cite{OD}
\begin{equation}
S=\frac{1}{16\pi G_5}\int
d^5x\sqrt{-g}(\mathcal{R}-\frac{1}{2}\partial_M\phi\partial^M\phi-V(\phi)-\frac{f(\phi)}{4}F_{MN}F^{MN}),\label{action}
\end{equation}
where $G_5$ is the five-dimensional (5D) Newton constant. $g$
stands for the determinant of the metric $g_{MN}$. $\mathcal{R}$
represents the Ricci scalar.  $\phi$ denotes the scalar which
induces the deformation away from conformality. $V(\phi)$ refers
to the potential which contains the cosmological constant and some
other terms. $F_{MN}$ and $f(\phi)$ are the field strength tensor
and gauge kinetic function, respectively.

For (\ref{action}), the equations of motion are
\begin{equation}
\mathcal{R}_{MN}-\frac{1}{2}\mathcal{R}g_{MN}=T_{MN},\label{mo}
\end{equation}
\begin{equation}
\nabla_M(f(\phi)F^{MN})=0,
\end{equation}
\begin{equation}
\nabla^M\nabla_M\phi=V^\prime(\phi)+\frac{f^\prime(\phi)}{4}F_{MN}F^{MN},
\end{equation}
with
\begin{equation}
T_{MN}=\frac{1}{2}[f(\phi)(F_{MA}F_N^A-\frac{1}{4}g_{MN}F_{AB}F^{AB})+(\partial_M\phi\partial_N\phi-\frac{1}{2}g_{MN}\partial_A\phi\partial^A\phi)-g_{MN}V(\phi)],\label{mo1}
\end{equation}
where $\nabla_M$ represents the Levi-Civita covariant derivative
related to $g_{MN}$.

Notice that with $\phi=0$, the AdS black hole arises as a solution
of (\ref{mo})-(\ref{mo1}). Next, by considering a non-vanishing
gauge-field component $A_0$ with the conditions
\begin{equation}
A_0(z=0)=\mu R,\qquad A_0(z_0=z_h)=0,
\end{equation}
and solving (\ref{mo})-(\ref{mo1}), one can obtain the AdS-RN
spacetime, whose metric is given by
\begin{equation}
ds^2=\frac{R^2}{z^2}(-f(z)dt^2+d\vec{x}^2+\frac{dz^2}{f(z)}),\label{metric}
\end{equation}
with
\begin{equation}
f(z)=1-(1+Q^2)\frac{z^4}{z_h^4}+Q^2\frac{z^6}{z_h^6}, \qquad Q=\mu
z_h/\sqrt{3}
\end{equation}
where $R$ is the AdS radius. $Q$ denotes the black hole charge,
constrained in $0\leq Q\leq\sqrt{2}$. $z$ stands for the 5th
coordinate with $z=z_h$ the horizon and $z=0$ the boundary. $\mu$
represents the chemical potential. Note that the chemical
potential implemented here is not the quark (or baryon) chemical
potential of QCD but a chemical potential corresponding to the
R-charge of $\mathcal{N}=4$ SYM. Nevertheless, it could serve as a
simple way of introducing finite density effects into the system.

Moreover, the temperature of the black hole is
\begin{equation}
T=\frac{1}{\pi z_h}(1-\frac{Q^2}{2}). \label{T}
\end{equation}

Explicitly, for given $(T,\mu)$ in the boundary theory, $(z_h,Q)$
of the bulk theory can be expressed as
\begin{equation}
z_h(T,\mu)=\frac{3\pi
T}{\mu^2}[\sqrt{1+\frac{2}{3\pi^2}(\frac{\mu}{T})^2}-1],
\end{equation}
\begin{equation}
Q(T,\mu)=\frac{\sqrt{3}\pi
T}{\mu}[\sqrt{1+\frac{2}{3\pi^2}(\frac{\mu}{T})^2}-1].
\end{equation}

To emulate confinement in the boundary theory at vanishing
temperature, one can introduce a warp factor to the background,
similar to \cite{OA1}. Then the metric of the SW$_{T,\mu}$ model
is given by \cite{PCO}
\begin{equation}
ds^2=\frac{R^2}{z^2}h(z)(-f(z)dt^2+d\vec{x}^2+\frac{dz^2}{f(z)}),
\qquad h(z)=e^{c^2z^2}, \label{metric1}
\end{equation}
where $c$ is deformation parameter (related to the confining
scale) which has the dimension of energy. Here we will not focus
on a specific model, but rather study the behavior of the drag
force in a class of models parametrized by $c$. So we will make
$c$ dimensionless by normalizing it to fixed temperatures and
express other quantities in units of $c$. In Ref \cite{HLL}, it
was shown that the range of $0\leq c/T\leq2.5$ may be most
relevant for a comparison with QCD. We will use this range here.

To proceed, if one uses $r=R^2/z$ as the radial coordinate, then
metric (\ref{metric1}) turns into
\begin{equation}
ds^2=\frac{r^2h(r)}{R^2}(-f(r)dt^2+d\vec{x}^2)+\frac{R^2h(r)}{r^2f(r)}dr^2,\label{metric2}
\end{equation}
with
\begin{equation}
f(r)=1-(1+Q^2)(\frac{r_h}{r})^4+Q^2(\frac{r_h}{r})^6,\qquad
h(r)=e^{\frac{c^2R^4}{r^2}},
\end{equation}
now $T=\frac{r_h}{\pi R^2}(1-\frac{Q^2}{2})$,
$\mu=\frac{\sqrt{3}Qr_h}{R^2}$. The horizon and boundary are
$r=r_h$ and $r=\infty$, respectively. It should be noticed that
metric (\ref{metric1}) and metric (\ref{metric2}) are equivalent
but with different coordinate systems.

\section{drag force}
In this section, we will follow the argument in \cite{GB,CP} to
analyze the drag force in the SW$_{T,\mu}$ model. The string
dynamic is described by the Nambu-Goto action, given by
\begin{equation}
S=-\frac{1}{2\pi\alpha^\prime}\int d\tau d\sigma\sqrt{-g},
\label{S}
\end{equation}
with $g$ the determinant of the induced metric and
\begin{equation}
g_{\alpha\beta}=g_{\mu\nu}\frac{\partial
X^\mu}{\partial\sigma^\alpha} \frac{\partial
X^\nu}{\partial\sigma^\beta},
\end{equation}
with $g_{\mu\nu}$ the brane metric and $X^\mu$ the target space
coordinates.

One considers a heavy quark moving with constant speed $v$ in one
direction, (e.g., the $x_1$ direction) and takes the gauge
\begin{equation}
t=\tau, \qquad x_1=vt+\xi(r),\qquad x_2=0,\qquad x_3=0,\qquad
r=\sigma,\label{par}
\end{equation}
then the induced metric can be obtained by plugging (\ref{par})
into (\ref{metric2}),
\begin{equation} g_{tt}=-\frac{r^2f(r)h(r)}{R^2}, \qquad
g_{xx}=\frac{r^2h(r)}{R^2},\qquad g_{rr}=\frac{R^2h(r)}{r^2f(r)}.
\end{equation}

Given that, the Lagrangian density reads
\begin{equation}
\mathcal
L=\sqrt{-g_{rr}g_{tt}-g_{rr}g_{xx}v^2-g_{xx}g_{tt}{\xi^\prime}^2}=h(r)\sqrt{[1-\frac{v^2}{f(r)}+\frac{r^4f(r)}{R^4}{\xi^\prime}^2]},
\end{equation}
with $\xi^\prime=d\xi/d\sigma$. As the action does not depend on
$\xi$ explicitly, the energy-momentum current $\Pi_\xi$ is a
constant,
\begin{equation}
\Pi_\xi=\frac{\partial\mathcal L }{\partial
\xi^\prime}=\xi^\prime\frac{h(r)r^4f(r)/R^4}{\sqrt{1-\frac{v^2}{f(r)}+\frac{r^4f(r)}{R^4}{\xi^\prime}^2}}=constant\label{lag},
\end{equation}
yielding
\begin{equation}
\xi^\prime=\frac{\Pi_\xi
R^2}{r^2f(r)}\sqrt{\frac{f(r)-v^2}{r^4h^2(r)f(r)/R^4-{\Pi_\xi}^2}},\label{xi1}
\end{equation}
where $\xi^\prime$ can be regarded as the configuration of string
tail.

Near the horizon $r_h$, the denominator and numerator (inside the
square root of Eq.(\ref{xi1})) are negative for small $r$ and
positive for large $r$. Also, the ratio should be positive. Under
these conditions, one gets that the numerator and denominator
should change sigh at the same point (namely $r_c$). For the
numerator, one gets
\begin{equation}
f(r_c)=v^2,
\end{equation}
results in
\begin{equation}
Q^2(\frac{r_h}{r_c})^6-(1+Q^2)(\frac{r_h}{r_c})^4+1-v^2=0,\label{rc}
\end{equation}
note that for given values of $Q$ and $v$, the analytic
expressions of $r_h/r_c$ can be determined from Eq.(\ref{rc}) but
these are cumbersome and not very illuminating. Here we will
mainly focus on the numerical results.

The denominator also changes sign at $r_c$, yielding
\begin{equation}
{\Pi_\xi}=\frac{r_c^2h(r_c)\sqrt{f(r_c)}}{R^2},
\end{equation}
with
\begin{equation}
h(r_c)=e^{\frac{c^2R^4}{r_c^2}}.
\end{equation}

On the other hand, the current density reads
\begin{equation}
\pi_x^r=-\frac{1}{2\pi\alpha^\prime}\xi^\prime\frac{g_{tt}g_{xx}}{-g},
\end{equation}
and the drag force reads
\begin{equation}
f=\frac{dp_1}{dt}=\sqrt{-g}\pi_x^r,
\end{equation}
where the minus indicates that the drag force is against the
direction of motion, as expected.

Next, applying the relations
\begin{equation}
\lambda=g_{YM}^2N_c=\frac{R^4}{{\alpha^\prime}^2},\qquad
T=\frac{r_h}{\pi R^2}(1-\frac{Q^2}{2}),
\end{equation}
one arrives at the drag force in the SW$_{T,\mu}$ model
\begin{equation}
f=-\frac{\pi
T^2\sqrt{\lambda}}{2}\frac{vh(r_c)}{(1-\frac{Q^2}{2})^2(\frac{r_h}{r_c})^2}.\label{drag}
\end{equation}

Notice that by taking $Q=0$ (or $\mu=0$, which gives
$(\frac{r_h}{r_c})^2=\sqrt{1-v^2}$) and $c=0$ (corresponding to
$h(r_c)=1$) in Eq.(\ref{drag}), the results of $\mathcal{N}=4$ SYM
\cite{GB,CP} are recovered.

\begin{figure}
\centering
\includegraphics[width=8cm]{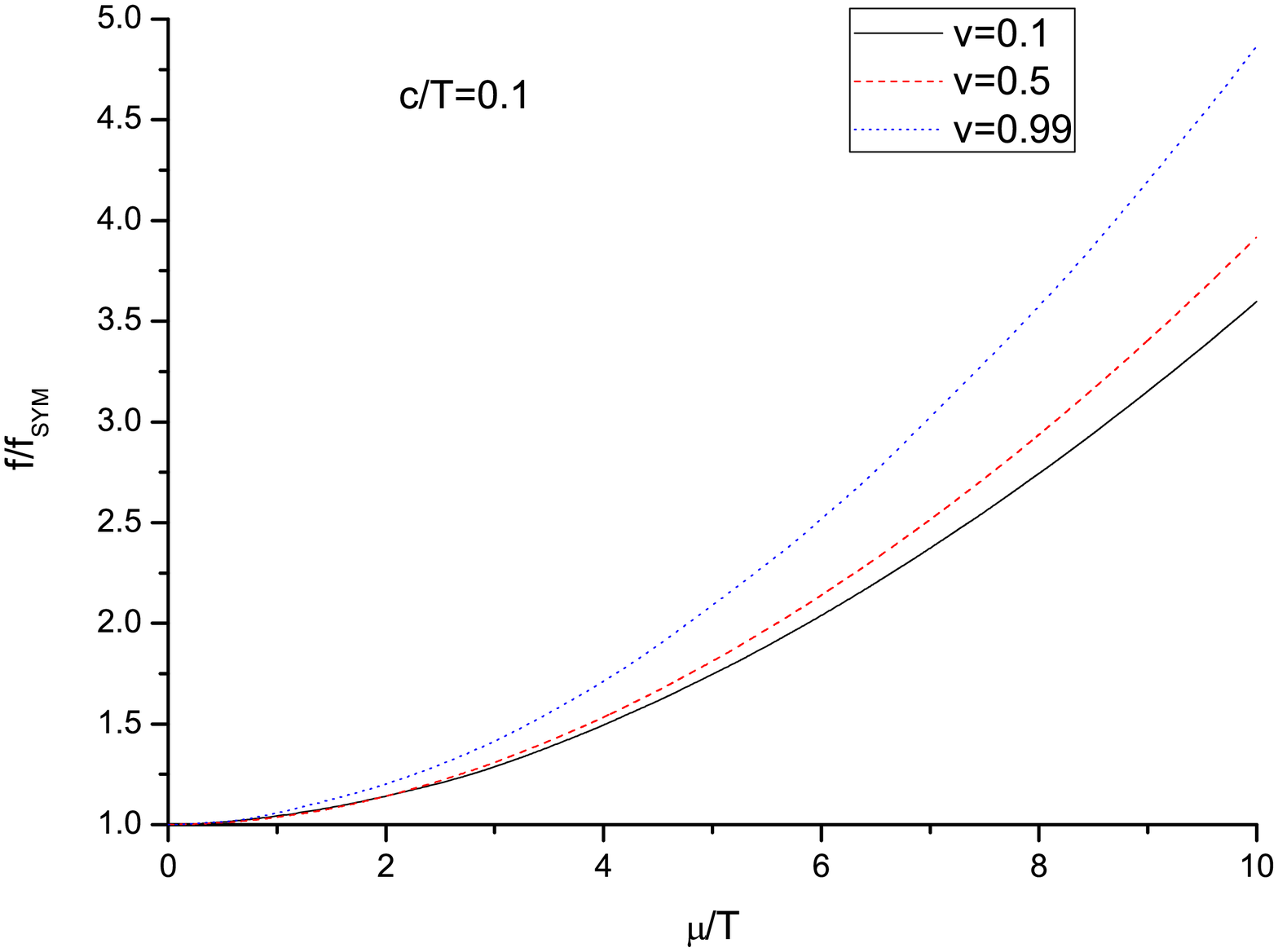}
\includegraphics[width=8cm]{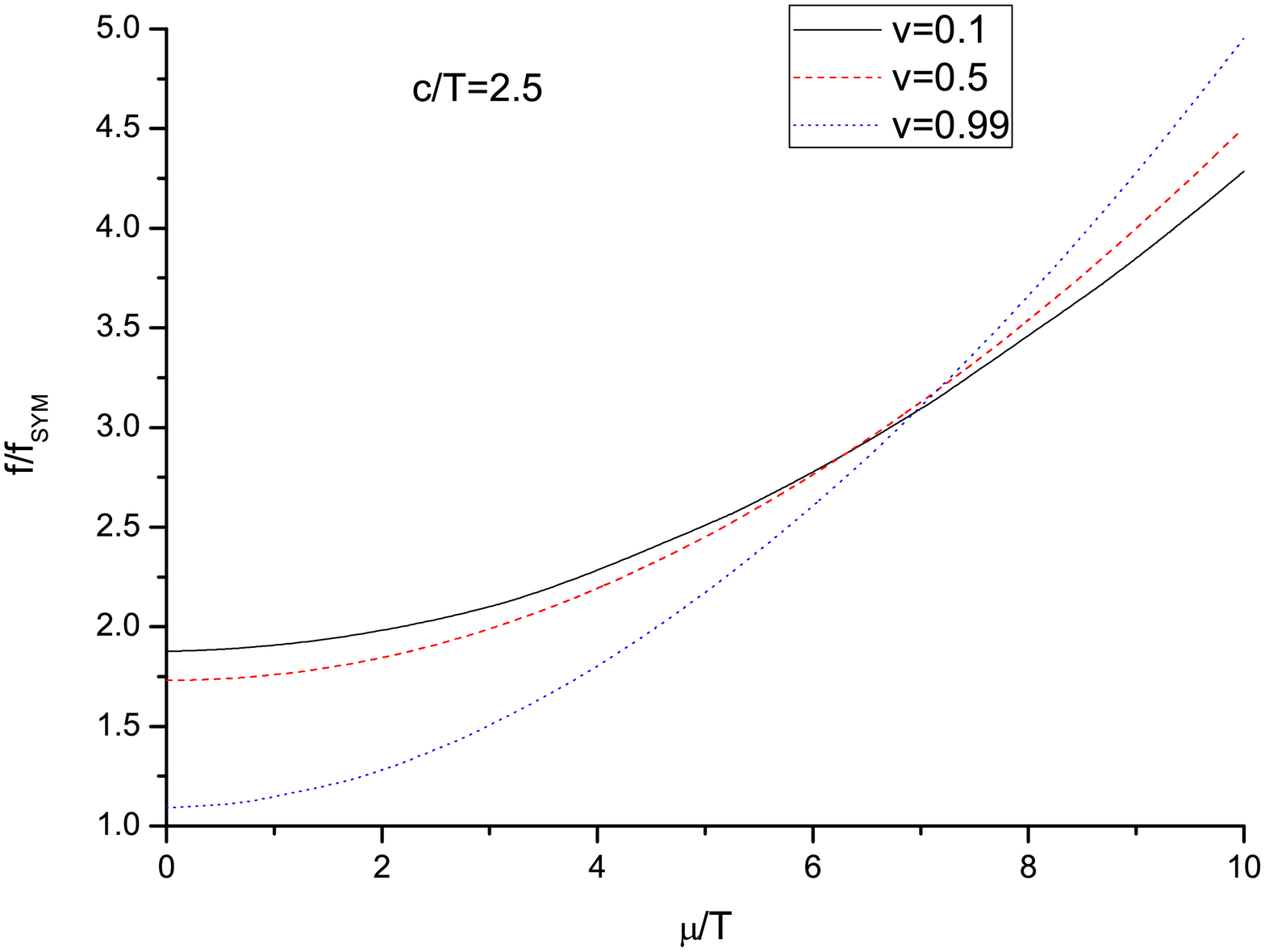}
\caption{$f/f_{SYM}$ versus $\mu/T$. Here we take the speed of
light as unit.}
\end{figure}

\begin{figure}
\centering
\includegraphics[width=8cm]{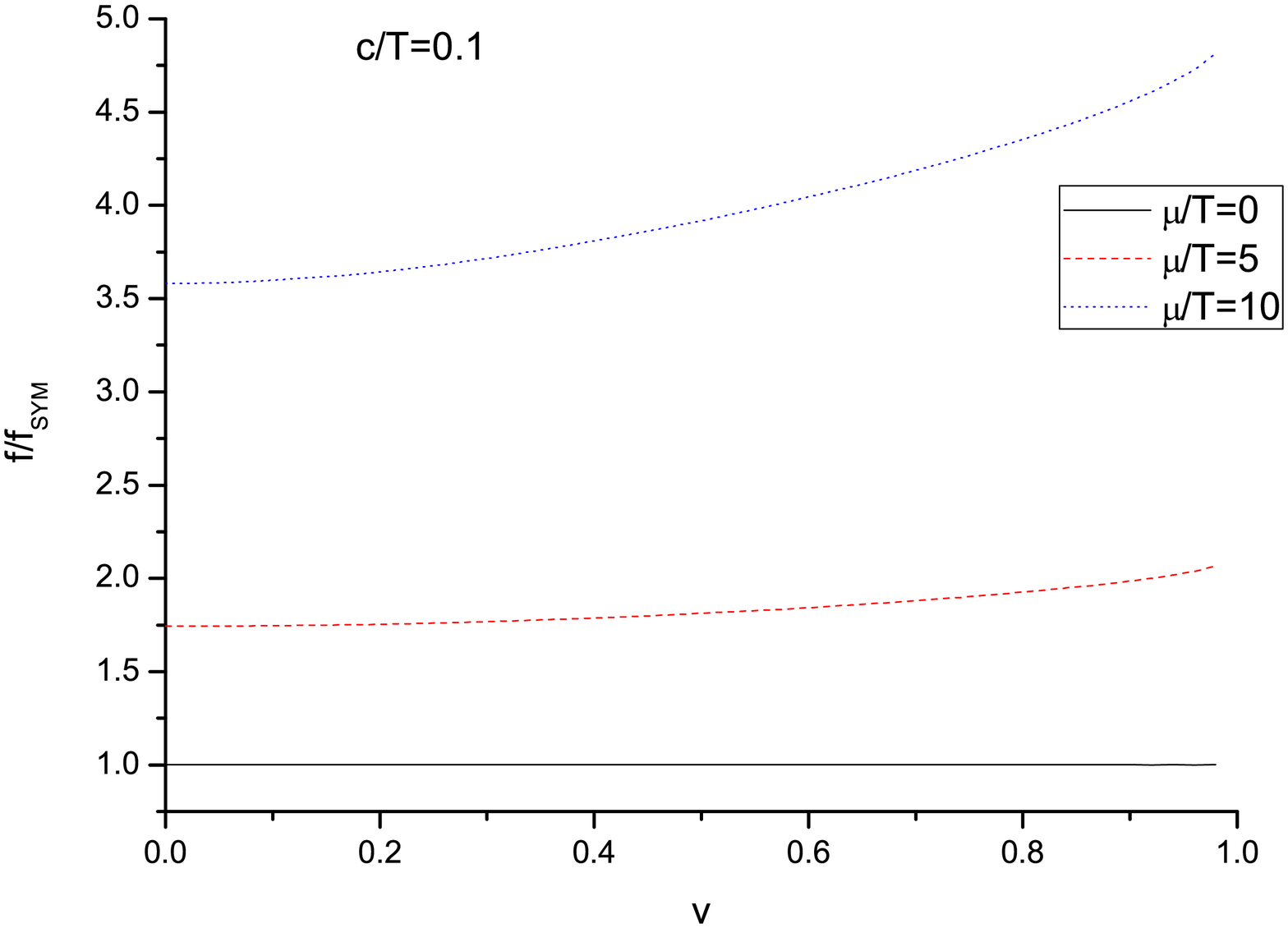}
\includegraphics[width=8cm]{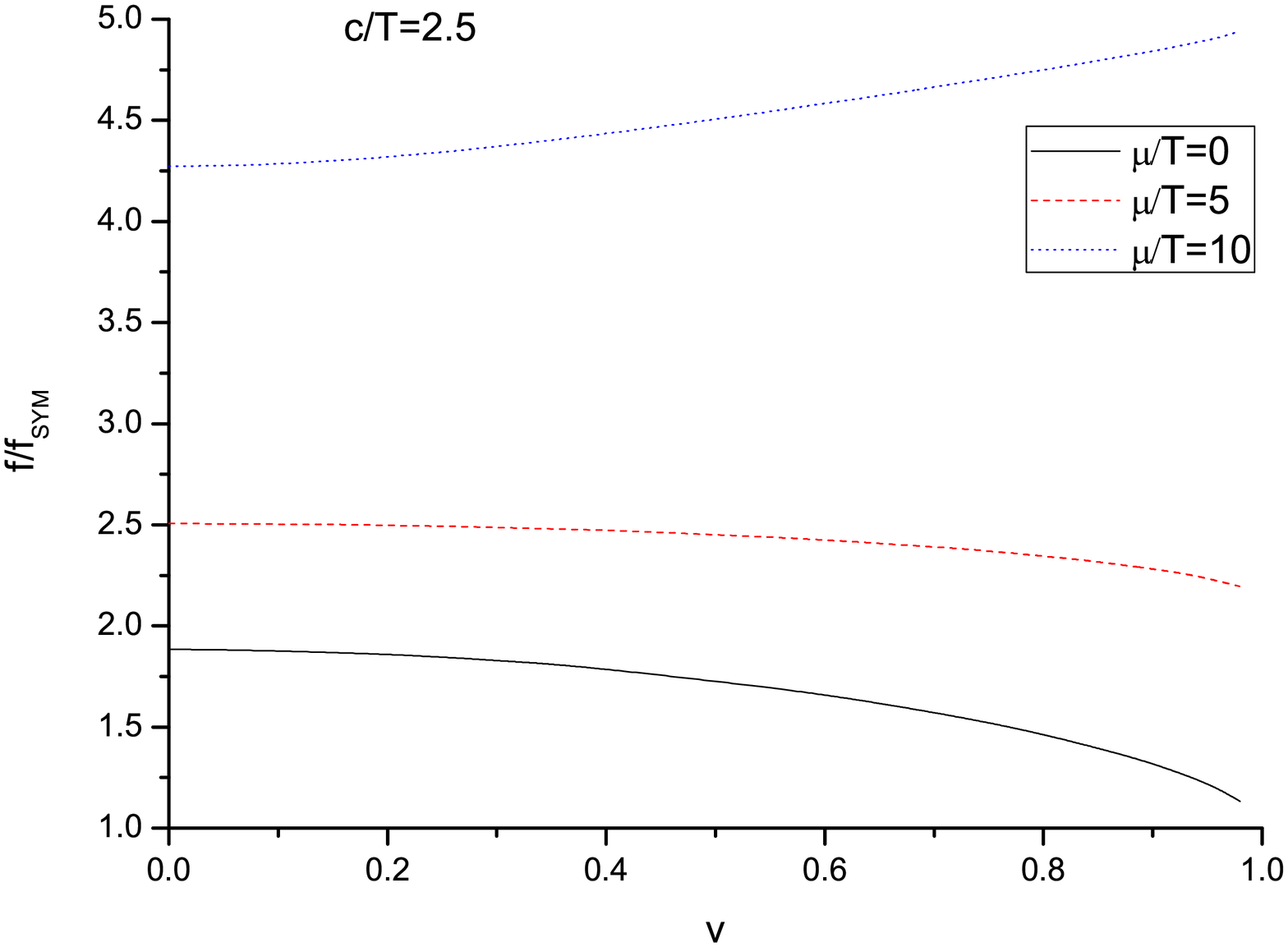}
\caption{$f/f_{SYM}$ versus $v$. Here we take the speed of light
as unit.}
\end{figure}

To proceed, we compare the drag force in the SW$_{T,\mu}$ model
with its counterpart of the $\mathcal{N}=4$ SYM as the following
\begin{equation}
\frac{f}{f_{SYM}}=\frac{h(r_c)\sqrt{1-v^2}}{(1-\frac{Q^2}{2})^2(\frac{r_h}{r_c})^2}.\label{drag1}
\end{equation}

In fig.1, we plot $f/f_{SYM}$ against $\mu/T$ for two fixed values
of $c/T$. The left panel is for $c/T=0.1$ and the right one is for
$c/T=2.5$. From these figures, one sees that the drag force in the
SW$_{T,\mu}$ model is larger than that of $\mathcal{N}=4$ SYM
plasma. Also, increases $\mu/T$ leads to increasing the drag
force. Namely, the chemical potential enhances the drag force,
consistently with \cite{ECA,SCH,LC}. In addition, by comparing the
two panels, one finds that increasing $c/T$ increases the drag
force. Thus, one concludes that the confining scale also enhances
the drag force, in accordance with \cite{ENA}.

In fig.2, we plot $f/f_{SYM}$ versus $v$ for various cases. From
the left panel (small $c/T$ case), one finds that increasing $v$
enhances $f$, especially this effect is more pronounced for large
$\mu/T$. But the right panel (large $c/T$ case) is different: for
small $\mu/T$, $f/f_{SYM}$ decreases with $v$ but for large
$\mu/T$ the situation reverses. Interestingly, a similar
non-monotone behavior appears in the studies of the drag force
with curvature-squared corrections \cite{KBF}.

Also, one could analyze the effect of $c$ and $\mu$ on the
viscosity. As we know, a stronger force implies a more strongly
coupled medium, closer to an ideal liquid. As the drag force in
the SW$_{T,\mu}$ model is larger than that of $\mathcal{N}=4$ SYM,
one could infer that the plasma is less viscous in the
SW$_{T,\mu}$ model than in $\mathcal{N}=4$ SYM plasma.

Before closing this section, we would like to mention that Ref
\cite{OA0} has studied the drag force in the soft wall model and
estimated the spatial string tension at finite $T$ and $\mu$
recently.

\section{diffusion coefficient}
In this section, we investigate the behavior of the diffusion
coefficient in the SW$_{T,\mu}$ model. First, we remember the
results of $\mathcal{N}=4$ SYM. The drag force is given by
\begin{equation}
f_{SYM}=\frac{dp_1}{dt}=-\frac{\pi
T^2\sqrt{\lambda}}{2}\frac{v}{\sqrt{1-v^2}}.\label{drag1}
\end{equation}

Assuming $p_1=mv/\sqrt{1-v^2}$, Eq.(\ref{drag1}) becomes
\begin{equation}
\frac{dp_1}{dt}=-\frac{\pi
T^2\sqrt{\lambda}}{2}\frac{p_1}{m}.\label{drag2}
\end{equation}

Integrating (\ref{drag2}), one obtains
\begin{equation}
p_1(t)=p_1(0)e^{-t/t_{SYM}},\qquad t_{SYM}=\frac{2m}{\pi
T^2\sqrt{\lambda}},
\end{equation}
where $t_{SYM}$ is the relaxation time of $\mathcal{N}=4$ SYM.

On the other hand, the diffusion coefficient is related to the
temperature, the heavy quark mass and the relaxation time as
\cite{GB,CP}
\begin{equation}
D=\frac{T}{m}t_D.\label{t}
\end{equation}

Using the above approach, one readily gets
\begin{equation}
D_{SYM}=\frac{2}{\pi T\sqrt{\lambda}}.
\end{equation}

Likewise, from (\ref{drag}) one can rewrite the drag force in the
SW$_{T,\mu}$ as
\begin{equation}
f=-\frac{\pi p_1
T^2\sqrt{\lambda}}{2m}\frac{\sqrt{(1-v^2)}h(r_c)}{(1-\frac{Q^2}{2})^2(\frac{r_h}{r_c})^2}.\label{drag1}
\end{equation}

In a similar manner, one arrives at the diffusion coefficient in
the SW$_{T,\mu}$ model
\begin{equation}
D=\frac{T}{m}t_D=-\frac{T}{m}\frac{p_1}{f}=\frac{(1-\frac{Q^2}{2})^2(\frac{r_h}{r_c})^2}{h(r_c)\sqrt{1-v^2}}\frac{2}{\pi
T\sqrt{\lambda}}.
\end{equation}

\begin{figure}
\centering
\includegraphics[width=8cm]{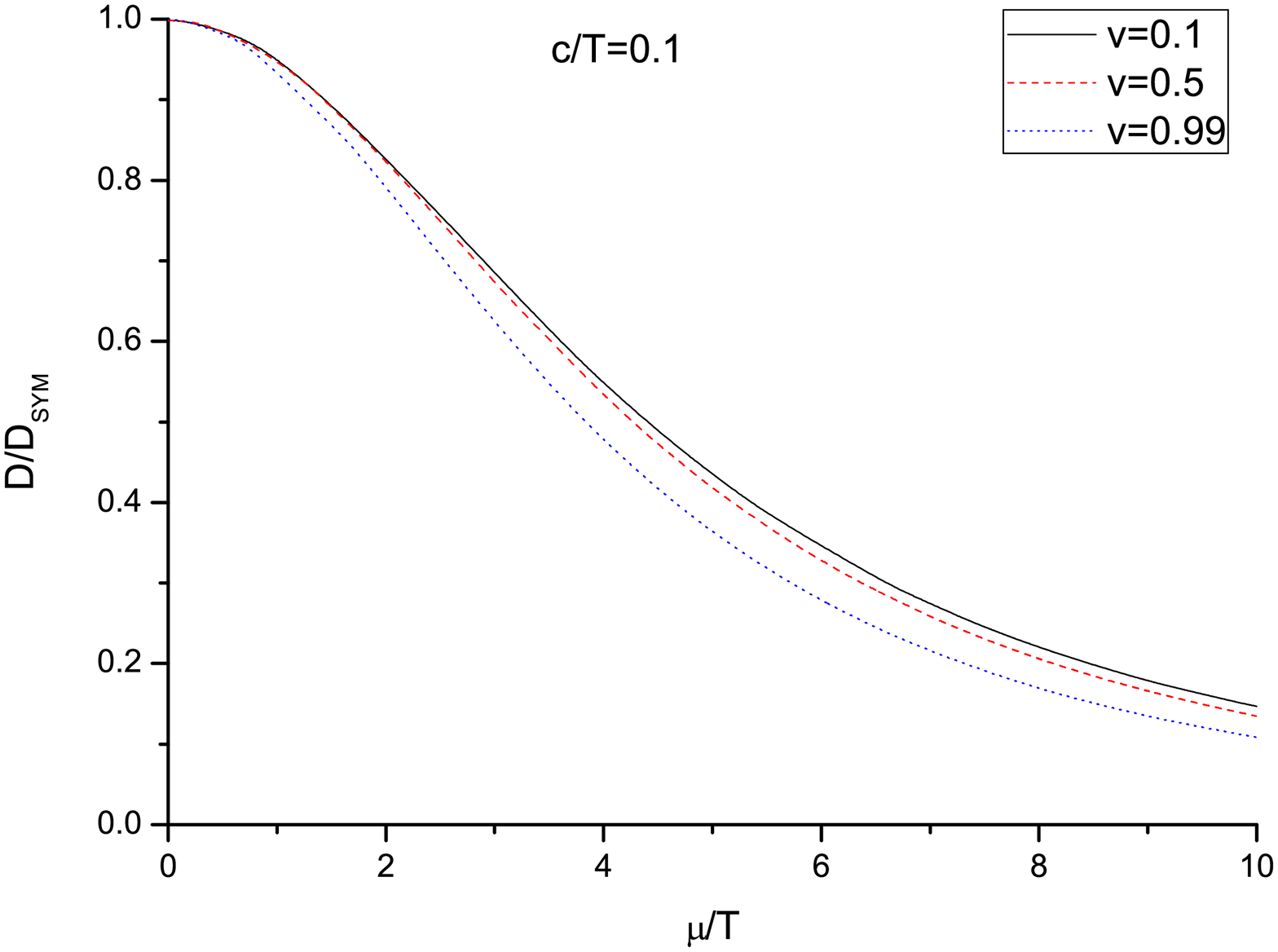}
\includegraphics[width=8cm]{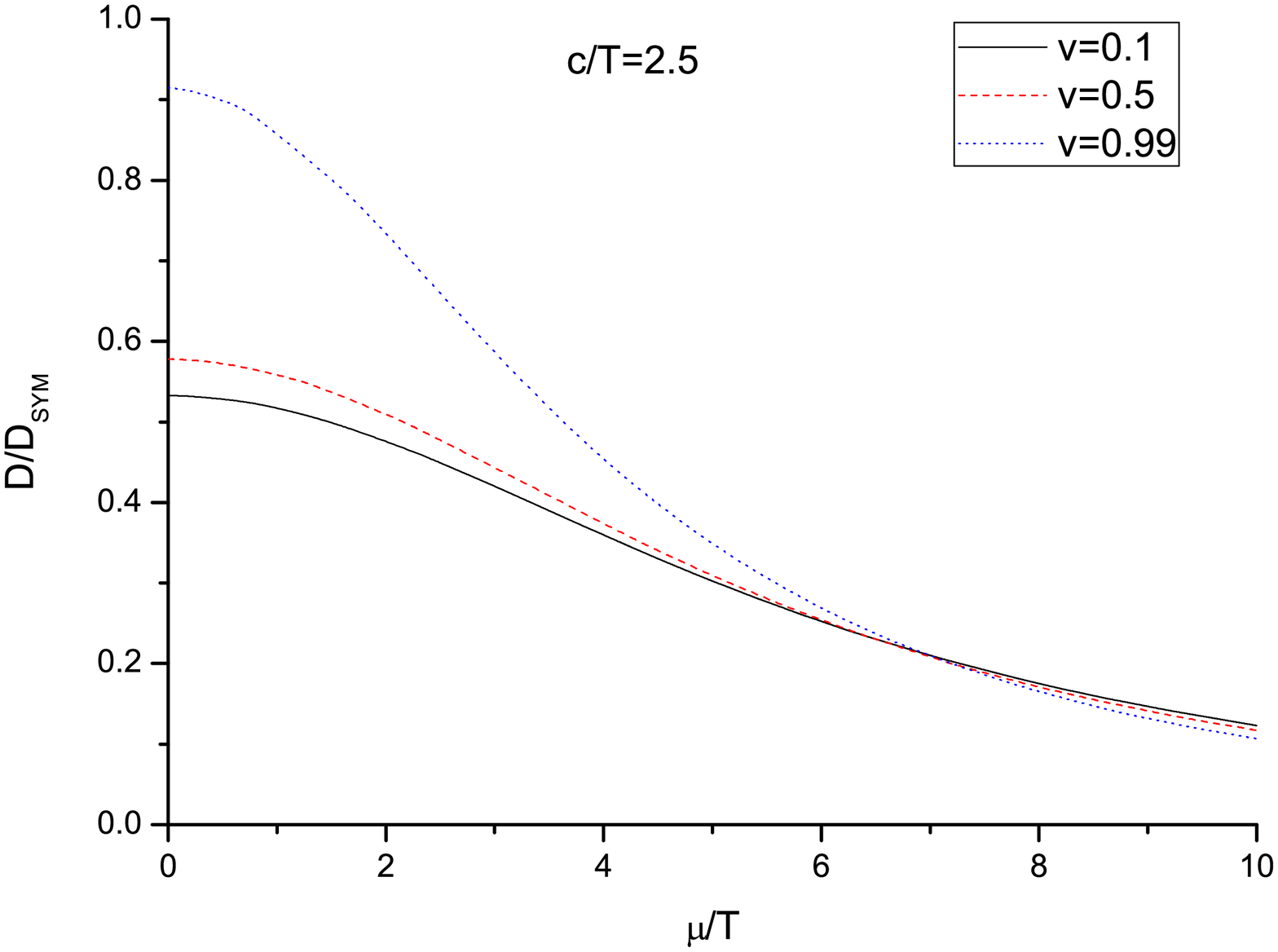}
\caption{$D/D_{SYM}$ versus $\mu/T$. Here we take the speed of
light as unit}
\end{figure}

In fig.3, we plot $D/D_{SYM}$ as a function of $\mu/T$ . One can
see that $\mu$ and $c$ both reduce the diffusion coefficient.

Also, the inclusion of $\mu$ and $c$ may influence the heavy quark
mass. As discussed in \cite{JCAS}, the relaxation time
$t_D=\frac{m}{T}D$ should be larger than the inverse temperature
\begin{equation}
t_D>>\frac{1}{T},
\end{equation}
yielding
\begin{equation}
m>>\frac{h(r_c)\sqrt{1-v^2}}{(1-\frac{Q^2}{2})^2(\frac{r_h}{r_c})^2}\frac{\pi
T\sqrt{\lambda}}{2}, \label{mass}
\end{equation}
one can see that $\mu$ and $c$ may enhance the heavy quark mass.

\section{Conclusion}
In this work, we studied the drag force and diffusion coefficient
in a soft wall model with finite temperature and chemical
potential. Our goal is to understand the interplay between the
presence of confining scale (so the plasma is not conformal) and
the chemical potential (the QGP is assumed to carry a finite,
albeit small, baryon number density) when estimating the drag
force of a heavy quark in a strongly coupled plasma. It is shown
that with fixed $c$ the presence of $\mu$ increases the drag force
and decreases the diffusion coefficient, in agreement with the
findings of \cite{ECA,SCH,LC}. Also, with fixed $\mu$ the
inclusion of $c$ increases the drag force as well, in accordance
with \cite{ENA}. Therefore, our results ($\mu$ and $c$ existed at
the same time) confirm the results of $\mu$ or $c$ stands alone.
On the other hand, the results show that the plasma is less
viscous in the SW$_{T,\mu}$ model than in $\mathcal{N}=4$ SYM.

However, it should be admitted that the SW$_{T,\mu}$ model
consider here has several drawbacks. First, it is not a consistent
model since it does not solve the Einstein equations. It would be
interesting to give such analysis in some consistent models, e.g.
\cite{ja,dl,RCR,SH,DL0} (generally, the metrics of those models
are only known numerically, so the calculates are quite
challenging. Also, in these models the dilaton field would be
non-trivial, so the coupling of the dilaton to the word-sheet
should be taken into account \cite{UG1,JNO,MM}). Moreover, the
SW$_{T,\mu}$ model may miss a part about the phase transition,
since Refs \cite{DL,RG2,FZ,FZ1} argued that there may be a first
order phase transition if one sets the warp factor to be $c z^2$
and solves the equation of motion self-consistently. It would be
significant to consider these effects.

Finally, it is also of interest to study the jet quenching
parameter \cite{HLI} and stopping distance \cite{PMC,SSG,PAR} in
the SW$_{T,\mu}$ model and compare with the results of this work.
We leave this as a further study.

\section{Acknowledgments}
We would like to extend our gratitude to Dr. Zi-qiang Zhang for
helpful and encouraging discussions. This work is supported by
Guizhou Provincial Key Laboratory Of Computational Nano-Material
Science under Grant No. 2018LCNS002.

%%%%%%%%%%%%%%%%%%%%%%%%%%%%%%%%%%%%%%%%

\end{document}